# PRED-CLASS: Cascading Neural Networks for generalized protein classification and genome-wide applications

**Claude Pasquier, Vassilis J. Promponas and Stavros J. Hamodrakas**

*Faculty of Biology, Department of Cell Biology and Biophysics,*
*University of Athens, Panepistimiopolis, Athens 15701, Greece*

## Abstract

A cascading system of hierarchical artificial neural networks is presented, for the generalized classification of proteins into four distinct classes: Transmembrane, Fibrous, Globular and 'Mixed', from information solely encoded in their amino acid sequences. This system, named PRED-CLASS, is a direct descendant of the recently published PRED-TMR2 algorithm, which initially discriminates transmembrane (TM) from globular, water soluble proteins with considerable success for several representative data sets. The architecture of the individual component networks is kept very simple, reducing the number of free parameters (network synaptic weights) for faster training, improved generalization and avoiding overfitting the data. Capturing information from as little as 50 protein sequences spread along the 4 target classes (6 TM, 10 Fibrous, 13 Globular and 17 Mixed), PRED-CLASS was able to obtain 371 correct predictions out of a set of 387 proteins (success rate ~96%) unambiguously assigned into one of the target classes. Application of PRED-CLASS to several test sets and complete proteomes of several organisms, demonstrates that such a method could serve as a valuable tool in the annotation of genomic ORFs with no functional assignment or as a preliminary step in fold recognition and 'ab initio' structure prediction methods. Detailed results obtained on various data sets, completed genomes, along with a web sever running the PRED-CLASS algorithm can be accessed over the World Wide Web at the URL: http://o2.biol.uoa.gr/PRED-CLASS.

# Introduction

Prediction of protein tertiary structure from information contained in amino acid sequence remains a challenging problem in structural molecular biology, despite the fact that great progress has been achieved during the last few years [1,2,3]. Over the last three decades, several computational methods have been developed for the prediction of 1D structural features of proteins from their sequence alone. For example, secondary structure prediction schemes aim to propose (sometimes with noteworthy success, [4,5,6]) the probable locations of secondary structure elements. In many cases, they are reported to be a fundamental initial stage (or a refining final stage) for 'ab initio' calculations [7], fold recognition [8,9] or homology modeling techniques [10]. Machine learning techniques have been a common practice to mine the information 'hidden' in the vast amount of protein sequences resulting from completed and ongoing Genome Projects, combined with available experimental functional or structural knowledge or information. In particular, different types of artificial neural network predictors have often served as a powerful tool for this achievement [11,12,13,14]. However, algorithms to predict generalized topological features of protein molecular structures could prove to be useful tools as a preliminary step in protein structure prediction and/or functional class determination.

We have recently published a simple neural network based classification scheme (PRED-TMR2, [15]) to distinguish between transmembrane and globular, water soluble proteins, integrated with a previously reported method (PRED-TMR, [16]) for fast and accurate detection of transmembrane segments. We now extend the capabilities of the system by further classifying non-transmembrane proteins into three classes: fibrous (e.g. collagen, elastin), globular (e.g. various types of enzymes) and a last group of proteins composed of both fibrous and globular domains, mentioned hereinafter as 'mixed' proteins (e.g. several intermediate filament proteins).

In this work, we illustrate the learning procedure followed (neural network training), as well as results obtained on several sets of known proteins for an estimation of the classification error rate. The method's predictive power combined with reasonable time performance on currently available complete proteome sets (for instance the recently sequenced *Drosophila melanogaster* genome, [17]) indicates that PRED-CLASS could be a precious source of information in automatic or manual annotation of genomic 'orphan' ORFs (ORFans, [18]). The generalized classification obtained by the method suggests that PRED-CLASS could be useful as a starting point in fold recognition or 'ab initio' structure prediction methods, while, in combination with comparative studies on completed genomic sequences, it could give further insight in the evolution of protein structure and function.

# Results

**Results on an evaluation test set of 387 proteins**

In order to assess the performance of the system, several tests were applied. As described in previous work [15], NN1 demonstrated a perfect performance on a set of 101 non-homologous transmembrane proteins (101 correct predictions) and a very good performance in a subset of PDB_SELECT composed of 995 globular proteins (97.7% correct assignments to the globular class). We have created a new test set to contain a total of 387 well-characterized protein sequences from all 4 target classes (147 membrane, 73 fibrous, 55 globular and 112 mixed), as described in the Materials and Methods section, to evaluate the performance of the new integrated system.

|  |  | PREDICTED | | | | | |
|---|---|---|---|---|---|---|---|
|  |  | TM | FIBR | GLOB | MIX | Total Obs. | SEL (%) |
| OBSERVED | TM | 139 | 0 | 8 | 0 | 147 | 94.5 |
| | FIBR | 1 | 72 | 0 | 0 | 73 | 98.6 |
| | GLOB | 0 | 1 | 54 | 0 | 55 | 98.2 |
| | MIX | 3 | 3 | 0 | 106 | 112 | 94.6 |
| | Total Predicted | 143 | 76 | 62 | 106 | 387 | |
| | SENS(%) | 97.2 | 94.7 | 87.1 | 100.0 | | |

**Table 1.** System Performance on a test set of 387 protein sequences. SEL=selectivity, SENS=sensitivity, see equations (1), (2).

The overall performance of the classification scheme is approximately 96% (371 correct assignments), and a summary of the results obtained by PRED-CLASS is presented in Table 1. A first measure of the method's success is its selectivity (SEL) for each particular class $C$ (TM, FIBR, GLOB, MIX):

$$SEL_c = 100 * t_{pc} / (t_{pc} + f_{nc}) \qquad (1)$$

Selectivity represents the percentage of correct assignments in each class (true positive hits, $t_{pc}$) compared to the total members of the class (true positives plus false negatives, $t_{pc}+f_{nc}$), where $f_{nc}$ is the number of proteins belonging in class $C$, but erroneously classified in another class. A higher selectivity of 98.6% was obtained for the fibrous class and a lowest 94.5% for the transmembrane one, yielding a mean selectivity of 96.5%. It should be taken into consideration that the first classifier, which decides weather a protein sequence corresponds to a transmembrane protein or not, has been trained on as little as 11 sequences.

Sensitivity (SENS) of classification for each class $C$ is also a useful evaluation criterion:

$$SENS_c = 100 * t_{pc} / (t_{pc} + f_{pc}) \qquad (2)$$

that is the percentage of correct assignments in each class compared to the total number of assignments in this class (true positives plus false positives, $t_{pc}+f_{pc}$), where $f_{pc}$ is the number of erroneous assignments in class $C$. Along these lines, classification in the mixed class was 100% sensitive, yielding no false predictions in the mixed class, whereas a lowest sensitivity was obtained for the globular class (87.1%), with a mean sensitivity value of 94.7%. This result, however, could be rather misleading, since the globular class apparently is the most abundant in the protein universe, whereas in our test set this class of proteins is somehow under-represented and a more sensitive performance of the globular classifier should be expected. Additionally, since no mixed proteins were used as negative examples in the training phase of the first neural network, it seems sensible that 3 out of the 4 false assignments in the transmembrane class come from mixed proteins. A maximal collection of proteins in the fibrous, globular and mixed classes, removing redundancy like in the work reported by [19], would certainly serve as a valuable set for training and evaluating classification systems like PRED-CLASS.

4**Case by case study of erroneous classifications**

Wrong classifications, in this evaluation phase, have been examined manually in order to gain some better understanding of the method's performance. Half of the 16 erroneous predictions have been for 8 TM proteins classified in the globular class (SwissProt [20] IDs: COXK_BOVIN, COXD_BOVIN, TONB_ECOLI, VS10_ROTBN, GEF_ECOLI, FDOH_ECOLI, BCS1_YEAST, NNTM_BOVIN). Half of them (COXK_BOVIN, COXD_BOVIN, BCS1_YEAST, NNTM_BOVIN) are bound to the mitochondrion membrane, whereas 2 of them (TONB_ECOLI, GEF_ECOLI) contain a signal membrane-anchored N-terminal sequence.

Another erroneous classification occurred for a fibrous collagen type IV precursor (CA44_HUMAN) possessing a signal N-terminal sequence ranging from residue 1 to 38, incorrectly classified as TM. Careful inspection of the output of the first NN, reveals that the segment responsible for the wrong classification is located exactly in this region. After removing this signal peptide from the sequence it is correctly classified in the fibrous class.

A total number of 6 false predictions for mixed proteins was obtained, 3 of them being type II cytoskeletal keratins (K2CA_HUMAN, K2CB_HUMAN, K2CF_HUMAN) misclassified in the transmembrane class. These sequences are almost identical (more than 98% pairwise identity) and the false classification is due to a long segment rich in glycine and hydrophobic residues. The rest of the 3 wrongly predicted mixed proteins (K1CJ_BOVIN, K2M2_SHEEP, NFH_MOUSE) were assigned to the fibrous class.

A subtle case was the only globular protein that was classified in the fibrous class: Ferredoxin. Its characteristic cysteine-rich iron-binding domains exhibit a periodicity of approximately 3.4 residues and when 3 cysteines are artificially 'mutated' into another residue type ferredoxin is correctly classified into the fibrous class, indicating that a probable reason for this wrong assignment is both the overall composition and the detected periodic patterns.

**Results in the context of the SCOP classification of protein domains**

A rather qualitative test of the method was performed against the sequences of protein domains as classified in the SCOP database [21] choosing as the most appropriate resource for this task the ASTRAL [22] subset (see Materials and Methods), which is freely accessible via the Internet. The chosen threshold of sequence similarity in this data set, ensures that, on one hand, most of the redundancy (in terms of sequence similarity) is removed, while on the other hand, several sequence representatives exist in most of the classification levels. This data set (shortly SCOPlt40pc) contains 2619 sequences of protein domains while the whole SCOP version 1.48 contained 21828 protein domains. The distribution of PRED-CLASS predictions along the seven classes of the first level of SCOP hierarchy is presented in Table 2. Although most sequences in this test set correspond to globular water-soluble protein domains, there are several points to make without considering further details:

a) 2198 out of the 2286 protein domains (~96%) classified in the all-alpha, all-beta, alpha/beta and alpha+beta classes in the SCOP classification scheme, were predicted to belong to the Globular class. Only 43 of them (1.9%) were predicted to be transmembrane, 29 (1.3%) fibrous and 16 mixed (0.7%).

b) Membrane or cell surface protein domains are predicted to belong either to the transmembrane or the globular class (possible for membrane anchored protein domains)

c) Most of the domains classified by PRED-CLASS in the fibrous class correspond to 'small' protein domains, which are usually domains rich in disulfide bonds. No small proteins have been classified in the transmembrane or the mixed class.

| SPECIES | TM | FIBR | GLOB | MIX |
|---|---|---|---|---|
| **BACTERIAL** | | | | |
| *Aquifex aeolicus* | 21.9 | 0.1 | 72.8 | 5.2 |
| *Bacillus subtilis* | 26.1 | 0.2 | 68.9 | 4.7 |
| *Borrelia burgdorferi* | 25.6 | 0.2 | 70.5 | 3.7 |
| *Campylobacter jejuni* | 25.4 | 0.1 | 70.9 | 3.6 |
| *Chlamydia muridarum* | 26.7 | 0.3 | 67.2 | 5.7 |
| *Chlamydia pneumoniae* | 29.6 | 0.5 | 65.2 | 4.7 |
| *Chlamydia trachomatis* | 25.3 | 0.1 | 69.1 | 5.5 |
| *Deinococcus radiodurans* | 20.2 | 3.0 | 73.3 | 3.5 |
| *Escherichia coli* | 23.8 | 0.5 | 72.7 | 3.0 |
| *Haemophilus influenzae* | 21.3 | 0.2 | 74.6 | 3.8 |
| *Helicobacter pylori* | 22.6 | 0.1 | 71.5 | 5.7 |
| *Helicobacter pylori J99* | 23.2 | 0.3 | 71.0 | 5.4 |
| *Mycobacterium tuberculosis* | 21.3 | 4.6 | 71.9 | 2.1 |
| *Mycoplasma genitalium* | 27.1 | 0.0 | 68.7 | 4.1 |
| *Mycoplasma pneumoniae* | 22.8 | 0.0 | 72.0 | 5.1 |
| *Rickettsia prowazekii* | 29.7 | 0.1 | 66.8 | 3.3 |
| *Synecocystis sp. (strain PCC 6803)* | 26.7 | 0.5 | 69.4 | 3.4 |
| *Thermotoga maritima* | 23.8 | 0.3 | 70.4 | 5.5 |
| *Treponema pallidum* | 24.5 | 1.4 | 68.5 | 5.6 |
| *Ureaplasma parvum* | 24.4 | 0.2 | 73.1 | 2.3 |
| *Xylella fastidiosa* | 19.9 | 1.1 | 76.7 | 2.3 |
| **Mean** | **24.4** | **0.6** | **70.7** | **4.2** |
| **Standard deviation** | **2.7** | **1.1** | **2.7** | **1.2** |
| | | | | |
| **ARCHAEAL** | | | | |
| *Aeropyrum pernix* | 18.3 | 18.2 | 61.2 | 2.3 |
| *Arcaeoglobus fulgidus* | 21.3 | 0.2 | 75.2 | 3.2 |
| *Methanobacterium thermoautotrophicum* | 20.9 | 0.3 | 73.6 | 5.1 |
| *Methanococcus jannaschii* | 20.5 | 0.1 | 76.7 | 2.7 |
| *Pyrococcus abyssi* | 23.3 | 0.3 | 72.8 | 3.6 |
| *Pyrococcus horikoshii* | 27.4 | 2.0 | 67.3 | 3.2 |
| **Mean** | **21.9** | **3.5** | **71.1** | **3.3** |
| **Standard deviation** | **2.8** | **6.6** | **5.3** | **0.9** |
| | | | | |
| **EUKARYOTIC** | | | | |
| *Caenorhabditis elegans* | 38.2 | 0.8 | 17.1 | 43.8 |
| *Drosophiila melanogaster* | 24.6 | 3.8 | 53.0 | 18.6 |
| *Saccharomyces cerevisiae* | 28.7 | 0.5 | 55.2 | 15.6 |
| **Mean** | **30.5** | **1.7** | **41.8** | **26.0** |
| **Standard deviation** | **5.7** | **1.5** | **17.5** | **12.6** |
| | | | | |
| **Overall Mean** | **24.5** | **1.3** | **67.9** | **6.2** |
| **Overall Standard deviation** | **3.8** | **3.3** | **10.8** | **7.8** |

**Table 3.** PRED-CLASS results on 30 complete proteome sets. The percent of proteins predicted to belong in the transmembrane, fibrous, globular, mixed class respectively are listed.

A thorough analysis of the PRED-CLASS predictions against all SCOP levels of hierarchy is currently in progress, paying much attention in cases where PRED-CLASS predictions are not consistent within the same fold, family or even superfamily.

|                          | TM | FIBR | GLOB | MIX | TOTAL |
|--------------------------|----|------|------|-----|-------|
| **all-alpha**            | 15 | 9    | 447  | 13  | 448   |
| **all-beta**             | 1  | 12   | 588  | 1   | 602   |
| **alpha/beta**           | 16 | 4    | 610  | 2   | 632   |
| **alpha+beta**           | 11 | 4    | 553  | 0   | 568   |
| **multidomain**          | 3  | 0    | 44   | 1   | 48    |
| **membrane and cell surface** | 28 | 0 | 27   | 0   | 55    |
| **small**                | 0  | 43   | 187  | 0   | 230   |
| **Total**                | 74 | 72   | 2456 | 17  | 2619  |

**Table 2.** Distribution of PRED-CLASS predictions along the seven classes in the first level of SCOP hierarchy based on the SCOPlt40pc set.

**Genome-wide application of PRED-CLASS**

With the great impact of genomic sequences into the public databases, the main goal of all efforts to devise reliable predictions, on aspects of protein structure and function, is the tractability of the developed methods to handle huge data sets with efficiency, with respect to computational time and resources. Our method's time performance is rather reasonable, while no external parameters need to be defined by the end user, which makes automatic genome-wide predictions with PRED-CLASS possible. For example, the recently sequenced genome of the fruit fly *Drosophila melanogaster* [17], composed of 13604 protein sequences, was analyzed overnight on a Silicon Graphics O2 Workstation with 128Mbytes of main memory and one 300MHz R5000 processor.

Thirty completed non-redundant proteome sets have been obtained from the 'Proteome Analysis' web site at the European Bioinformatics Institute (EBI) server (http://www.ebi.ac.uk/proteomes). The organisms whose proteomes have been analyzed ranged from simple archaea and bacteria to complex eukaryotic multicellular organisms, i.e. *Drosophila melanogaster*, and the results are summarized in Table 3. A significant proportion of these sequences does not show clear sequence similarity to proteins of known structure or function and consequently remain uncharacterized.

Even if no detailed annotation can be obtained for these 'unique' protein sequences, it is very interesting to answer the following questions: 'What is the (approximate) expected frequency of proteins in the transmembrane, fibrous, globular and mixed class in a newly sequenced genome? How are these frequencies correlated with the taxonomy of the studied organism and/or specific environment factors or cellular processes?' Previous studies raised the question whether the proportion of membrane proteins encoded in a genome is correlated with its size or not [23,24] with contrasting conclusions.

According to the predictions obtained by PRED-CLASS, even when averaging the frequencies of each class on proteomes of the same kingdom (Figures 1a, 1b, 1c), the standard deviations calculated are rather high (see Table 3) in order to support a generalized hypothesis. This apparently leads to the conclusion that there is substantial variation even within the same domain of life and further phylogenetic relations as well as organism specific information should be considered before suggesting any hypothesis. For example, if we examine the frequency of predicted transmembrane proteins in bacteria, where frequency values range between 19.9% (*Xylella fastidiosa*) and 29.7% (*Rickettsia prowazekii*), in accordance to [22], the mean and median of the bacterial distribution are equal (both 24.4%) and close to the mean frequency of the complete set of proteomes (24.5%), whereas the 'extreme' values lie approximately two standard deviations from the mean, indicating a normal-like distribution.

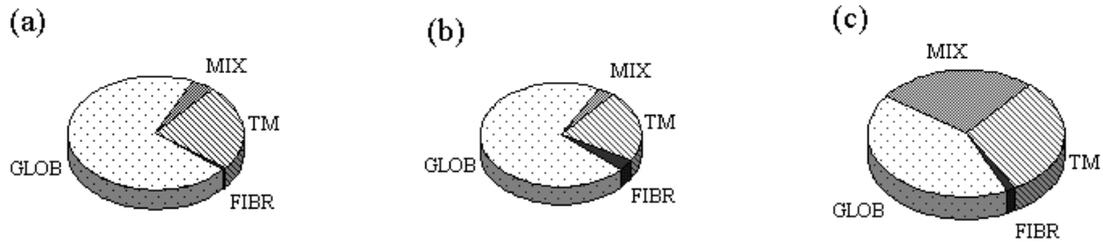

**Figure 1.** (a) Mean proportions of proteins encoded in 21 complete bacterial genomes classified into transmembrane, fibrous, globular and mixed classes as predicted by PRED-CLASS. (b) Mean proportions of proteins encoded in 6 complete archaeal genomes classified into transmembrane, fibrous, globular and mixed classes as predicted by PRED-CLASS. (c) Mean proportions of proteins encoded in 3 complete eukaryotic genomes classified into transmembrane, fibrous, globular and mixed classes as predicted by PRED-CLASS.

Most of the proteomes of archaeal and eukaryotic organisms studied in this work, fall between the limits imposed by the bacterial distribution for transmembrane proteins, with the only sound exception *Caenorhabditis elegans (C.elegans)*, where about 38% of the genes is predicted to encode for transmembrane proteins. This extremely high value cannot be counted as an artifact of our method as it is in agreement with previous works [23,24] following different strategies for this characterization. Another strange finding considering *C.elegans* is the fairly large number of proteins predicted to belong to the mixed class (43.8%) combined with a vary small incidence of predicted globular proteins (17.1%). These values are the maximum and minimum frequencies in the respective classes among all genomes analyzed in this study. Apparently, a detailed study of the *C.elegans* genome is necessary in order to explain these findings, combining several types of computer-based predictions with experimental knowledge.

## Discussion

The classification of protein structure and function is a major goal in structural molecular biology, aiming at understanding the principles that govern the folding procedure, when 'linear' amino acid chains fold to a three dimensional structure, adapting a preferable shape for their desired function. Here, we have demonstrated PRED-CLASS, a robust system of simple artificial neural networks for generalized classification of proteins using information encoded in single amino acid sequences. Although protein sequence information is submitted to the system in machine-friendly numerical representations, the underlying principles are based on well-known attributes of protein structure.

The high overall prediction accuracy (~96%), indicates that PRED-CLASS could be pertinent for single-sequence or genome-wide analysis, either as a stand alone application or as a part of more elaborate computational analysis. PRED-CLASS could be utilized as an assisting tool in genome functional annotation projects, where some potential functions for a protein sequence become more possible while others are excluded after a correct prediction in the one class or the other. Such generalized predictions could prove to be valuable in primary stages of threading methods or 'ab initio' protein structure prediction. In the former, predictions could reduce the number of folds possibly compatible with an examined sequence, while in the later case structural constraints (e.g. existence of coiled-coil regions in fibrous or mixed proteins) are indicated, drastically improving execution time and performance of such approaches.

As a stand alone method, PRED-CLASS has been utilized to analyze several complete proteomes from all the domains of life (archaea, eubacteria, eukaryots). It is possible that there exists a tendency in eukaryotic complete genomes to code a larger proportion of transmembrane proteins, in agreement with previously reported work [23]. This finding could be explained by the invention of new protein functions through evolutionary mechanisms to accommodate the need of multicellular organisms for communication with the cell environment. However, due to the insufficient complete genome information in this domain of life and lack of the appropriate experimental evidence it may be too early to make such an assumption.

Plans for future work in the improvement and further extension of the proposed system are already in progress, resulting from some weaknesses of the method (described in the Results section) and the emerging need for divergent accurate predictors. Transmembrane protein detection could apparently gain in sensitivity by filtering in a preprocessing stage secretory signal peptides, while sensitivity could be improved by re-training NN1 with mitochondrial transmembrane proteins as well.

A feature useful to the end users would be to be presented with a reliability index accompanying each prediction, so that the 'quality' of individual predictions could be known before deciding further analysis steps. A main problem with this task is that large data sets of proteins with reliable characterization into all 4 classes should become available.

Although the time performance of the system's current version is adequate, a parallel implementation of PRED-CLASS would make genome-wide predictions faster by an estimated factor of 1.5-2. A novel fold-class prediction method for globular, water-soluble proteins is currently in the final development and testing stages and is intended to be integrated in the PRED_CLASS system.

## Materials and Methods

### Information gathering

A set of eleven proteins (6 transmembrane, 2 fibrous, 3 globular) with known structural characteristics has been used for the training of the first network, as described in detail elsewhere [15]. Another set of 40 proteins (10 fibrous, 13 globular and 17 mixed, see Table 4) has been selected for the learning process of the newly employed neural networks. This makes a total of only 50 protein sequences, considering that the sequence with corresponding SProt ID ELS_CHICK has been used in the training of all three networks.

Another set of 387 protein sequences reliably assigned to the four target classes (147 membrane, 73 fibrous, 55 globular, 112 mixed) served as a test data set to evaluate the predictive power of the system.

The non-redundant set of 148 integral membrane proteins recently compiled by Möeller and coworkers [19] served as an ideal representative set of well characterized prediction targets of the transmembrane class. For classification purposes, no detailed information concerning the location or the orientation of membrane spanning segments is necessary. One protein used in the training process of the first component network (SProt ID: LECH_HUMAN) was present in this set and was therefore removed. This set, which is larger than the initial test set on which PRED-TMR2 was tested upon, has been used to gain a more realistic approximation of the performance of the first network classifier.

The 55 representative globular proteins are a subset of the 65 well known globular proteins collected by Levitt and Greer [25]: they are typical globular proteins varying in sequence and functional characteristics. All 10 sequences present in the training set have been removed as well.

| **FIBROUS** | **GLOBULAR** | **MIXED** |
|---|---|---|
| CA16_HUMAN | ADHA_UROHA | ABP2_HUMAN |
| CA1E_HUMAN | AZU1_METJ | AINX_RAT |
| CCC4 * | C550_THIPA | DESM_CHICK |
| CH18_DROVI | CATA_YEAST | FILS_BOVIN |
| CH19_DROGR | CTR2_VESCR | GFAP_HUMAN |
| ELS_CHICK | DYRA_CITFR | IF3T_TORCA |
| FBOH_BOMMO | HBA1_ARCGA | IFE_BRALA |
| KR2A_SHEEP | IGJ_HUMAN | ION3_CARAU |
| KRB2_SHEEP | LYC1_CAPHI | LAM1_HUMAN |
| TPM1_CHICK | RNPH_BACSU | NEST_RAT |
| | RUB1_PSEOL | NF60_LOLPE |
| | THTR_AZOVI | NFM_MOUSE |
| | TRYP_ASTFL | PERI_RAT |
| | | PLST_CARAU |
| | | TANA_XENLA |
| | | VIM4_XENLA |
| | | XNIF_XENLA |

**Table 4.** SwissProt identification codes for sequences composing the training set used for NN2, NN3. CCC4 has not been deposited in SwissProt. It is a structural protein found in the egg-shell of the fruit fly *Ceratitis capitata* [30]

A set of 130 sequences belonging to the mixed class were extracted from SwissProt (release 35), seeking for entries containing all of the following keywords in their 'FT' fields: 'HEAD', 'ROD', 'TAIL', which correspond to typical domain definitions in known mixed structures. All 17 sequences used for the training phase, as well as an incomplete sequence (SProt ID IFEA_HELPO), have also been removed.

The collection of fibrous proteins was based on expert knowledge deposited in the existing literature and is composed of typical representatives of this class, for example, different collagen types, non-cytoskeletal keratins, elastins, fibrous chorion proteins, all collected from SwissProt release 35.

We have to note that retrieving representative sets of non-homologous fibrous and mixed protein sequences was not an easy task, because of the low number of proteins in these classes deposited in the public databases. In addition, no safe automatic way exists to extract fibrous protein sequences from public databases, as there are no specific and selective keywords. As far as the mixed class is concerned, performing a search keywords 'HEAD', 'ROD', and 'TAIL' within the 'FT' fields in SwissProt release 39, yielded just 55 more protein sequences. For this reason, we have decided to contain all protein sequences from SwissProt release 35 that, to our knowledge, could fit into these two classes so that the test could be as general as possible, not taking possible sequence similarity into account. As more data become available, further testing on larger data sets should be performed.

We also considered that another useful test would be to screen the classification results in the context of the SCOP classification scheme of protein domains [21]. On these grounds, we chose the ASTRAL (Brenner et al., 2000) subset of SCOP (version 1.48) sequences with less than 40% pairwise similarity. This set is composed of 2619 sequences of protein domains classified (in the first level of SCOP-version 1.48-hierarchy) in seven classes: 484 all alpha, 602 all beta, 632 a/b, 568 a+b, 48 multidomain, 55 membrane and cell surface proteins and peptides, 230 small proteins.

Protein sequence data collected for the training and evaluation phases of the PRED-CLASS system are available over the WWW at the URL: http:/o2.biol.uoa.gr/PRED-CLASS.

**System Architecture and Component Neural Network topology**

The overall classification system consists of three successive multilayer feedforward (acyclic) artificial neural networks (NNs, Figure 2), each one with a single hidden layer, where the computation takes place. NN1 has been previously described [15]. Some common features shared by all three NNs are:

i) 'Full Connectivity', as every node in each network layer is connected to every other node in the adjacent forward layer.

ii) Small number (2,3,3 respectively) of nodes in the hidden layer, those that are responsible for the actual learning process carried out by each component network.

iii) The 'activation function' on each node is a non-linear sigmoidal logistic function [26] of the weighted sum of all synaptic weights (plus a constant bias, not shown in Figure 2).

iv) Output values are normalized in the 0-1 range.

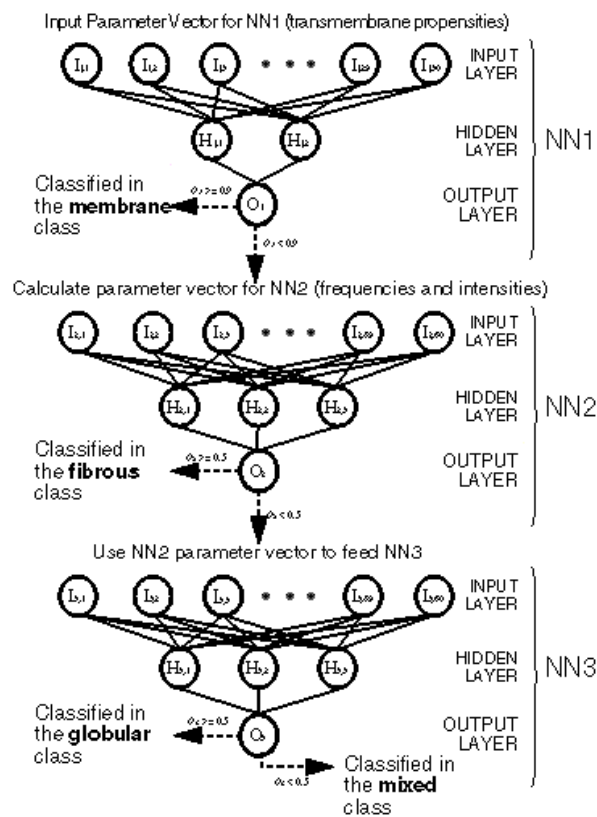

**Figure 2.** PRED-CLASS arhitecture: Individual component neural networks and their layered structure. Input type for each subsystem, network connectivities, information flow and decision scheme for the output layer of each NN are indicated.

In addition, NN2 and NN3 have an input layer of 60 nodes, which means that 60 parameters for each sequence are required to feed the network during the learning or classification operation. Before the learning process, all network synaptic weights are initialized to small random values.

Using error backpropagation [27] as the learning algorithm for each NN, two passes of computation are required during the training stage: A forward pass, where all synaptic weights remain unaltered and network signals are computed on each neuron in an hierarchical order by the activation function. The value on the output node is then compared with the desired network response (ideally 1 for positive classification, 0 for rejection) which corresponds to the already known class of the parameter vector presented to the network. An error signal is computed as the difference of

4
the obtained and the desired network output, which switches-on the backward pass, starting at the output layer, propagating error signals to adjacent backward layers of neurons, performing an adjustment of synaptic weights. The target of this phase is to minimize the error value, following a gradient descent strategy. Several sequential iterations, with randomized order of presentation of training examples might be necessary, until the network is considered to have converged, stabilizing the error rate below a desired threshold, ideally leading to a global minimum.

By keeping the number of synaptic weights as small as possible the training process is lead faster to convergence, since less free parameters have to be readjusted for optimization in each backpropagation cycle. At the same time the number of training examples required to achieve good generalization (small fraction of classification errors) can be reduced, as it is reported to be proportional to the number of free parameters [28].

In the prediction phase, just like the forward pass in learning, network weights are globaly fixed (those obtained after the convergence of the training process) and the NN is presented with an 'unknown' example for classification. In the same hierarchical manner, the input signal propagates once in the forward direction and the output value constitutes the network's decision based on the already studied training examples.

**Training Parameters**

In our study, those proteins classified as non-membrane by NN1 are candidates for classification by NN2. After a careful inspection of the sequential and structural characteristics of several fibrous, globular and mixed proteins (data not shown) the following 60 input parameters where chosen as an appropriate input to NN2:

(i) 30 values corresponding to the composition of the examined sequence in all 20 residue types and 10 different groupings of residues sharing common structural and/or physicochemical properties. The amino acid groupings used in this study were the following: AVLIFWDEQMHK (α-helix formers), VLIFWYTCQM (β-sheet formers), GPDNSCKWYQTRE (β-turn formers), CVILMFYWAP (hydrophobic), DEHKRSTNQ (polar), HRKDE (charged), HRK (positively charged), DE (negatively charged), HFWY (aromatic) and VLIA (aliphatic).

(ii) 30 values corresponding to the highest intensity for periodicities detected for each residue or group type by a Fast Fourier Transform (FFT) algorithm [29]. The implementation of the FFT algorithm within this method is applied with the default parameters, encoding protein sequences in a numerical string of 0s and 1s to note the absence or presence respectively of any desired residue type in a specific position in the examined sequence. Higher intensities for a particular amino acid type (or group) suggest the existence of an underlying periodic pattern in which this residue type is involved.

These two types of parameters clearly reflect patterns of composition as well as relative distributions of amino acid types along sequences and repetitive elements. Although such an approach might seem rather simplified, NNs are capable of capturing subtle patterns in available data and succeed to recognize weak underlying signals 'buried' in the examined data.

The output neuron of NN2 is considered activated (or 'fired' in the NN terminology) when its value ($O_2$) is $O_2 \geq 0.5$, indicating a positive case of a fibrous protein. In the opposite case ($O_2 < 0.5$) the sequence is further examined by NN3, which has exactly the same topology and accepts identical parameters to NN2. A result $O_3 \geq 0.5$ in NN3's output node indicates a case of a globular protein, otherwise the protein is classified as mixed.

**Acknowledgements**

The authors acknowledge the support of the EEC-TMR 'GENEQUIZ' grant ERBFMRXCT960019.